\documentclass[natbib,authoryear]{sigtbd17-style}

{\makeatletter \gdef \@copyrightdata {\hspace{-2.5em}%
  \rlap{\textcolor{white}{\rule{2.5em}{1em}}}%
  \protect\namestack{Erik D. Demaine; Martin L. Demaine}}}

\usepackage[utf8]{inputenc}

\usepackage{graphicx}
\usepackage[scaled]{helvet} 
\usepackage[hyphens,spaces,obeyspaces]{url}
\usepackage{enumitem}      
\usepackage[colorlinks=true,allcolors=blue,breaklinks,draft=false,unicode]{hyperref}   
\newcommand{\doi}[1]{doi:~\href{http://dx.doi.org/#1}{\Hurl{#1}}}   

\usepackage{xcolor}
\definecolor{commentgreen}{RGB}{2,112,10}
\definecolor{ouridentifier}{RGB}{0,71,171} 
\definecolor{frenchplum}{RGB}{129,20,83}
\definecolor{stringorange}{RGB}{139,64,0} 
\definecolor{identifieryellow}{RGB}{82,81,0}
\usepackage{listings}
\usepackage{multicol} 

\def\topfraction{0.99}

\def\floatpagefraction{0.99}

\def\BibTeX{\textsc{Bib}\TeX}

\usepackage[final]{pdfcomment} 
\usepackage{accsupp}           
\usepackage{pgf}               

\newcommand\textopacity[2]{%
  \begin{pgfpicture}%
    \pgfsetfillopacity{#1}%
    \pgfpathmoveto{\pgfpointorigin}%
    \pgftext[base]{#2}%
  \end{pgfpicture}%
}

\makeatletter
\newcommand\textstack[3][;]{%
  \def\textstack@append##1{\expandafter\textstack@step##1#1\@eol}%
  \def\textstack@step##1#1##2\@eol{%
    \hbox{\textopacity{#2}{\ignorespaces ##1\unskip}}%
    \ifx\@eol##2\@eol\else
      \vskip-\baselineskip
      \textstack@step##2\@eol
    \fi}%
  \pdftooltip{%
    \BeginAccSupp{method=pdfstringdef,unicode,ActualText={#3}}%
      \vbox{\textstack@append{#3}}%
    \EndAccSupp{}%
  }{#3}%
}
\makeatother

\newcommand\namestack[2][0.666]{\textstack{#1}{#2}}

\input{stack-natbib}

\begin{document}

\title{Every Author as First Author%
  \thanks{Full source code of this paper is available at
    \url{https://github.com/edemaine/author-stack-paper}}}

%
%

\authorinfo{\namestack{Erik D. Demaine; Martin L. Demaine}}
{\makebox{Computer Science and Artificial Intelligence Laboratory} \\
\makebox{Massachusetts Institute of Technology}  \\
\makebox{Cambridge, MA 02139}}
{\namestack{edemaine;mdemaine}@mit.edu}

\maketitle

\begin{abstract}
  We propose a new standard for writing author names on papers
  and in bibliographies, which places
  \emph{every author as a first author --- superimposed}.
  This approach enables authors to write papers as true equals,
  without any advantage given to whoever's name
  happens to come first alphabetically (for example).
  We develop the technology for implementing this standard
  in \LaTeX, \BibTeX, and HTML;
  show several examples; and discuss further advantages.
\end{abstract}

\section{Introduction}

\paragraph{The problem.}

Authorship order makes for fraught debates in academia.
Perhaps the most common standard is to list authors in decreasing order
by significance of contribution.
But this quantity is usually difficult to measure,
and can lead to uncomfortable conversations, arguments, or even feuds.
For example, are an advisor's leadership and high-level ideas
more or less important than a student's technical solutions?
Is one author's technical work more or less important than
another author's writing of the paper?
Even if the answer to these questions are clear to you
(and equal among all authors), authors' contributions are rarely so clearcut.
And what if the contributions are roughly equal?

In some disciplines, the order has additional codified meanings.
For example, many natural sciences use the last author position to indicate the
research supervisor (principal investigator) whose lab housed the work.
But what if multiple people serve that role, as research becomes
increasingly collaborative?

\paragraph{Existing solutions.}

Some disciplines --- such as economics, mathematics, and
theoretical computer science --- discourage the idea of determining authorship
order by defaulting to an algorithmically determined order:
alphabetically increasing by surname.
In particular, the alphabetical standard is a central tenet of
\emph{supercollaborative} research \citep{supercollaboration},
where researchers brainstorm to solve problems as equals,
and everyone decides for themselves whether they contributed
enough to be an author.
The motivation is that it is difficult to compare contributions when
brainstorming, as failed ideas are often as important as successful ideas,
and ideas are often as important as technical work.
Agreeing to alphabetical authorship ahead of time guarantees that everyone
will be recognized for their contributions, without ever having to argue
about who contributed what.

The alphabetical approach avoids any uncomfortable conversations and arguments,
and works well when contributions are roughly equal or difficult to compare.
But in practice, we have occasionally seen authors who feel slighted
by being listed late despite having contributed significantly more than others
(e.g., having led the research and/or paper).

Other disciplines offer footnotes to clarify authorship order. For example,
multiple authors can be marked ``joint first author'' to indicate
equal contributions, or multiple authors can be marked as having
``jointly supervised'' the work.
In these cases, the authors in the same category
are normally listed alphabetically.
When all authors contributed roughly equally,
many papers include a footnote explaining that authorship order is alphabetical
(especially in publication venues where this is not the standard).
\emph{Nature} encourages including author contribution statements which
specify each author's exact contribution to the work.

\paragraph{Bias.}

A fundamental limitation to \emph{any} approach that lists the authors
in a fixed order arises when citing papers with several authors.
In the body of a paper (as opposed to the bibliography), it is most common
to write ``X et al.~[\#]''\ when referring to a paper [\#]
whose first author's surname is X.
In author--year styles such as APA, this is even built into the citation
itself, e.g., (X et al., 2023).
As a result, author X gets their name effectively promoted with every citation,
which is inconsistent with multiple or all authors being equal.

In our own writing, we try to avoid this practice, and instead
write all authors' surnames whenever citing a paper,
e.g., ``X, Y, and Z~[\#]''.
But this workaround becomes impractical for references with over a dozen
authors, such as some of our papers
\cite{ArithmeticGames_ISAAC2020,LessThanEdgeMatching_JIP}
or some papers in astronomy \cite{reverse-alphabetical} and
biology \cite{human-genome}.

Beyond listing authors when citing papers,
other biases arise from alphabetical ordering specifically.
Bibliography styles where papers are sorted alphabetically (such as ACM's)
cluster together papers with the same first author, further promoting people
who have alphabetically early surnames so are more likely to be first.
And of course fields whose standard is not alphabetical ordering
unfairly judge authorship order of papers that do.

These effects are collectively referred to as
\emph{alphabetical discrimination} \cite{alphabetical-discrimination}.
Several studies have explored this phenomenon, and find evidence that
people with alphabetically earlier surnames are more likely to succeed
academically.
The present authors have sometimes uncomfortably wondered whether they have
benefitted in this way, with surnames starting with ``D''.

To compensate for alphabetical discrimination,
several specific papers have explored alternate mechanisms
for deciding authorship order,
as documented in a footnote.
These mechanisms include competition via
25-game croquet series \cite{croquet},
2-day backgammon contest \cite{backgammon},
tennis match \cite{tennis},
basketball free throws \cite{free-throws},
arm wrestling \cite{arm-wrestling},
brownie bake-off \cite{brownie-bakeoff},
a game of chicken \cite{chicken}, or
rock paper scissors \cite{rock-paper-scissors};
by coin toss \cite{coin-toss},
dice roll \cite{dice},
the outcome of famous cricket games \cite{cricket},
currency exchange rate fluctuation \cite{currency-fluctuation}, or
dog treat consumption order \cite{dog};
or by authors' height \cite{height},
fertility \cite{fertility},
proximity to tenure \cite{tenure-proximity},
reverse alphabetical order \cite{reverse-alphabetical},
or degree of belief in the paper's thesis \cite{belief}.
Others have proposed games such as Russian roulette
(``publish and perish'') \cite{publish-and-perish}.
See the excellent surveys \cite{survey1,survey2,survey3}
and their comments.


\section{Solution}

Our proposed solution is a new standard for listing the authors of a paper:
instead of any ordered list, write all author names on top of each other.
For example, instead of ``Erik Demaine and Martin Demaine'' (alphabetical),
we write \namestack{Erik Demaine; Martin Demaine}.
In this way, we achieve the true ideal of an unordered set of equal authors,
where every author comes first.

In settings where authors did not contribute roughly equally, we can generalize
to writing each group of equal authors as an overlapping stack of names.
For example, when we want to distinguish multiple first authors,
multiple ``middle'' authors, and multiple supervising authors,
we can write three distinct groups of authors,
where each group is overlapping.

Our vision is that each set of names (as we often associate with each paper)
becomes recognizable as its own image.  Readers can then recognize repetitions
of the same paper without necessarily reading the individual names.
For example, compare two of our papers with overlapping but differing sets
of 13--15 authors (which came from a common supercollaborative open problem
session so share many names):
\citet{ArithmeticGames_ISAAC2020} vs.\ \citet{LessThanEdgeMatching_JIP}.
Each author stack has a distinct shape.
If we now repeat one --- \citet{ArithmeticGames_ISAAC2020} ---
you should be able to recognize which it is.

\subsection{Revealing the Names}

Of course, we want to give authors (equal) credit for their papers,
not remove all credit.
We have implemented two ways to reveal the actual names present
in an overlapping stack, when viewing a PDF file on a computer.

\begin{figure}[htbp]
  \centering
  \hrule

  \includegraphics[scale=0.5]{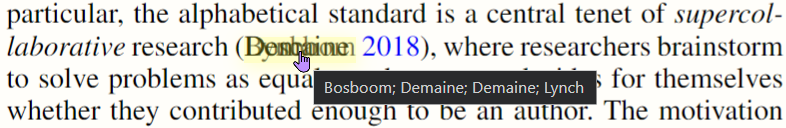}

  \hrule

  \includegraphics[scale=0.5]{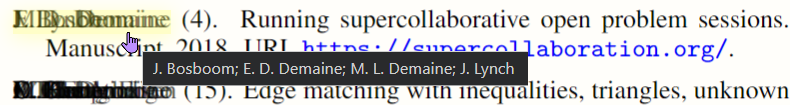}
  \caption{Hovering over a name stack reveals the authors in their original
    order, as they would normally be written in the citation (top)
    or bibliography (bottom).  Screenshots from
    VSCode's LaTeX Workshop Extension's internal PDF viewer
    \cite{latex-workshop-pdf}, which is based on PDF.js.}
  \label{fig:tooltip}
\end{figure}

First, hovering over the stacked names should pop up a tooltip with the
authors listed in their original order, as shown in
Figure~\ref{fig:tooltip}.
This feature works on many desktop PDF viewers (e.g., Acrobat, Evince, Firefox,
VSCode), but notably not Chrome%
\footnote{See \url{https://bugs.chromium.org/p/chromium/issues/detail?id=1122489}},
Edge, Safari, or MacOS Preview.
It also does not work on mobile devices we tested
(probably because they lack a natural notion of ``hovering'').
Because tooltips and links are both types of PDF annotations, they would
conflict with each other if used together, so we had to remove hyperlinks
from names; the years remain clickable links to the bibliography.

Second, copying and pasting the author stack (together with any surrounding
text) into another document should reveal the authors in their original order,
using PDF's accessibility feature.
This makes it easy to quote portions of a paper,
including any citations, with or without author stacks.
This feature is currently supported by only a few PDF viewers,
including Acrobat, Chrome, Edge, and Evince.

\subsection{Opacity}

To improve legibility, we write each name in semi-transparent ink
(currently, $2/3$ opacity).
As a result, where multiple characters overlap, the ink appears darker,
making the names more legible.  Compare:
\begin{itemize}
\item \namestack[1]{Erik Demaine; Martin Demaine} ~ (opacity $1$, no transparency)
\item \namestack[0.75]{Erik Demaine; Martin Demaine} ~ (opacity $3/4$)
\item \namestack[0.666]{Erik Demaine; Martin Demaine} ~ (opacity $2/3$)
\item \namestack[0.5]{Erik Demaine; Martin Demaine} ~ (opacity $1/2$)
\end{itemize}
In the future, we may consider using a different opacity depending on the
number of authors.  More extreme, we could use opacity to indicate the
relative contribution of each author (when they are unequal), or different
colors to represent different roles (such as supervisor).
But for now we like the uniformity of every name
appearing the same in all contexts.

\subsection{Space Savings}

An additional benefit of our solution is that long lists of names can be
written in far less space, and roughly the same amount of space for each paper.
For example, the 274 authors of the human genome project \cite{human-genome}
would normally take almost half a page to list:

\begin{quote}\sloppy
J.~C. Venter, M.~D. Adams, E.~W. Myers, P.~W. Li, R.~J. Mural, G.~G. Sutton,
  H.~O. Smith, M.~Yandell, C.~A. Evans, R.~A. Holt
% ...
\end{quote}

As a result, many papers that cite this paper do not list the entire
author list even in the bibliography, instead writing ``J. C. Venter et al.''
Our approach makes it easy to give the full author list \cite{human-genome},
giving credit to all authors, but without allocating that paper a
disproportionate amount of space in the bibliography.
Thus we help achieve fairness between cited papers, not just between authors
on each paper.

The space-savings property is also helpful for conferences and journals
with a hard limit on the number of pages including the bibliography.
Such limits are common among printed publications.
For such publications, the space savings also translates to a cost savings
for the publisher, and a reduction in trees needed to make paper ---
a helpful step toward resolving the climate crisis.

\section{Technology}

\subsection{\LaTeX}

We implemented \LaTeX\ macros for easily superimposing author names
into an ``overlay stack''.
A user can build such a stack from a semicolon-separated list of names
like so:

\begin{center}
  \lstinline|\namestack{Erik Demaine; Martin Demaine}|

  $\to$ ~
  \namestack{Erik Demaine; Martin Demaine}
\end{center}

Listing~\ref{lst:namestack} below shows the source code for this macro.
The main idea is to render each name (into a \TeX\ horizontal box),
and stack them vertically (in a \TeX\ vertical box)
with a negative one-line vertical space in between consecutive names.
The example above effectively becomes:

\begin{lstlisting}
\vbox{%
  \hbox{Erik Demaine}%
  \vskip-\baselineskip
  \hbox{Martin Demaine}%
}%
\end{lstlisting}

This main content gets wrapped in three additional components:
\begin{enumerate}
\item \lstinline|\textopacity{0.666}{...}| (implemented via \texttt{pgf.sty})
  to make each name semi-transparent at opacity $2/3$.
  The opacity can also be overridden via an optional argument, as in
  \lstinline|\namestack[0.9]{...}|.
\item \lstinline|\pdftooltip| (from \texttt{pdfcomment.sty})
  to add tooltips to the stack, with the original list of names.
\item \lstinline|\BeginAccSupp{...ActualText=...}...\EndAccSupp{}|
  (from \texttt{accsupp.sty}) to override copy/paste behavior for the stack.
\end{enumerate}

Listing~\ref{lst:natbib} shows additional code necessary to support
name/year bibliographies via \texttt{natbib.sty}.  Specifically, we
automatically wrap the ``name'' portion of a citation in
\lstinline|\namestack| (so it should be written as a semicolon-separated list).
We also need to disable hyperlinks on the name portion of each citation,
to avoid covering the tooltip with another PDF annotation.

\subsection{\BibTeX}

We modified the \texttt{abbrvnat.bst} \BibTeX\ bibliography style requested by
this proceedings to work with \lstinline|\namestack|;
refer to Listing~\ref{lst:bst} for the code.
First, we automatically wrap author/editor names within each bibliography entry
with \lstinline|\namestack{...}|.
We also append the number of names in a parenthetical;
this helps the reader know how many names are in each stack,
which we find particularly helpful in the context of this paper
for getting the feel for the newly introduced stacks.
Second, we output \texttt{natbib.sty}-compatible labels with semicolons
between all author names (instead of just the first author name followed by
``et al.''), so that the code in Listing~\ref{lst:natbib} constructs the
appropriate author stack.

\subsection{HTML}

\begin{figure}[htbp]
  \centering
  \includegraphics[scale=0.2]{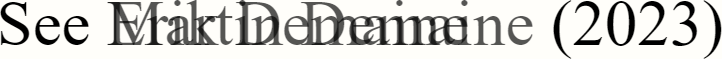}
  \caption{Rendering of HTML in Listing~\ref{lst:html}.
    Screenshot from Google Chrome 111 on Windows 11.}
  \label{fig:html}
\end{figure}

We also give a proof-of-concept that the same style of rendering is
possible in HTML.  Figure~\ref{fig:html} shows a rendering
of the HTML code in Listing~\ref{lst:html}.
CSS Grid makes implementing an overlapping stack quite easy:
by placing each name in the same grid cell,
the cell automatically grows to the maximum size of all names.

\begingroup
\lstset{
  language=HTML,
  alsoletter={-},
  otherkeywords={grid-row, grid-column, opacity, display, inline-grid},
  emph={stack, name}, 
  identifierstyle=,
  morecomment=[s]{/*}{*/}
}
\begin{lstlisting}[caption={HTML source code illustrating a name stack},label=lst:html]
<style>
.stack { display: inline-grid; }
.stack > .name {
  grid-row: 1; grid-column: 1;
  opacity: 0.666; /*new stacking context*/
}
</style>

See
<span class="stack">
  <span class="name">Erik Demaine</span>
  <span class="name">Martin Demaine</span>
</span>
(2023)
\end{lstlisting}
\endgroup

\section{Future Work}

One limitation of our new standard for listing names is that longer surnames
gain some bias.
For example, ``Olds'' (the second half of ``Mitchell-Olds'')
is clearly visible in \citet{currency-fluctuation}.
Will this lead to longer and longer academic surnames in the future?
Arguably, longer names already receive some bias without our system,
as they take up more relative space in the bibliography.
Yet in practice so far, most surnames seem to have a similar length.
The readability of a \emph{suffix} of a surname also does not seem to
advantage the actual name much.

\begin{figure}
  \centering
  \includegraphics[width=\linewidth]{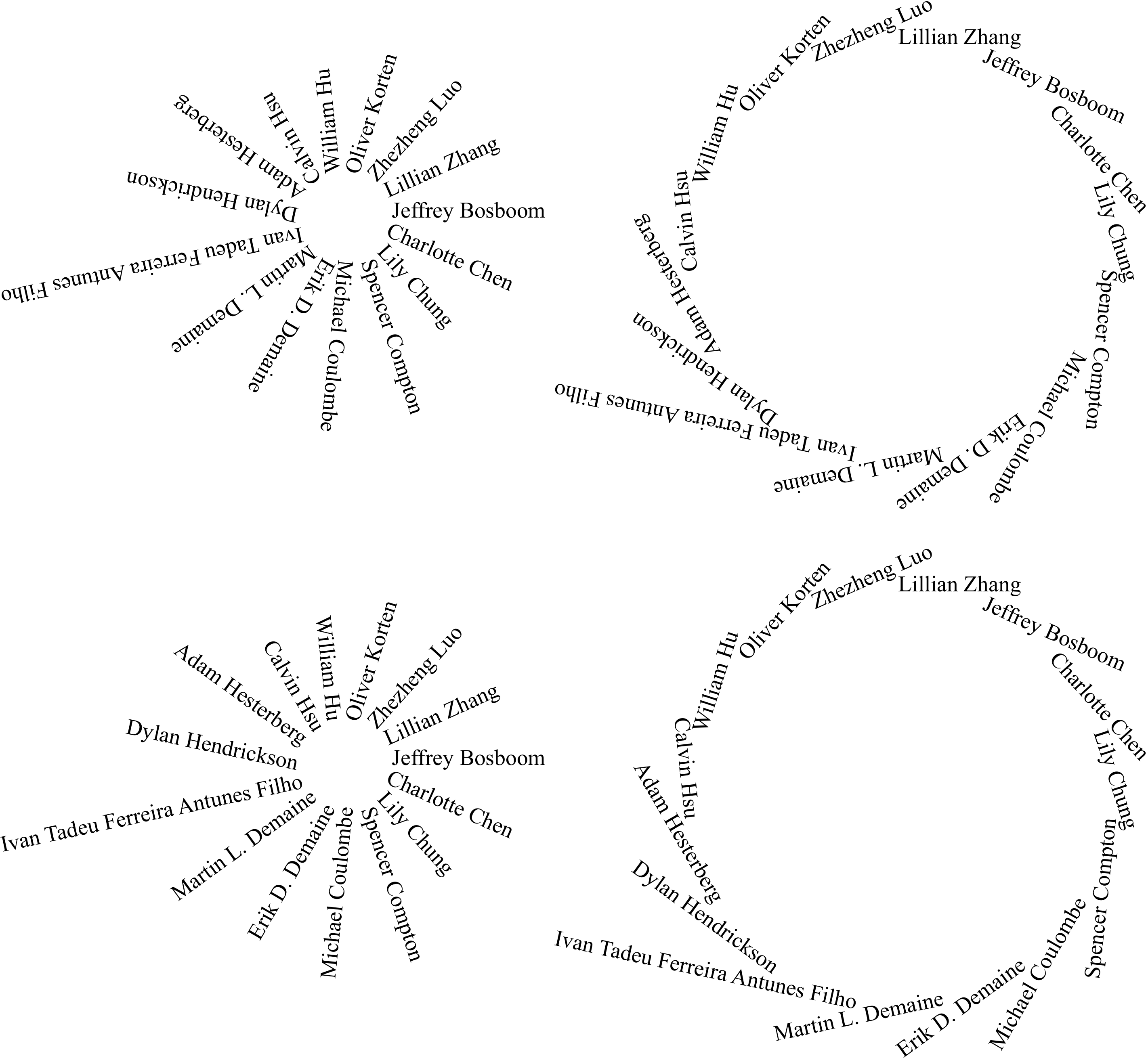}
  \caption{Circular arrangements of the authors of
    \citet{LessThanEdgeMatching_JIP}.  Drawn in Inkscape
    using Circular Align and Distribute, onto a circle of
    radius 50 (left) or 200 (right); followed by $90^\circ$ rotation
    (left); and rotating $180^\circ$ to make names upright (bottom).}
  \label{fig:circles}
\end{figure}

A final issue is that overlapping name stacks are not easy to read.
It may be possible to write names in a way that has no first name
but still makes all names clearly readable.
For example, a circle has no beginning or end, so arranging the names
in a circular pattern avoids arranging any author ``first''.
Figure~\ref{fig:circles} shows some initial experiments in this direction.
Related, traditional \emph{round-robin} documents \cite{round-robin}
are signed by authors in a circle to prevent identification of a ringleader
(such as mutineer sailors).
It remains unsolved how to fit such circular arrangements
in with the rest of a text document, which feels inherently sequential.
Circular arrangements also seem difficult to apply to small numbers of
authors such as $2$.

\begin{lstlisting}[float=*,caption={\LaTeX\ source code for \texttt{\textbackslash namestack}},label=lst:namestack]
%% \namestack for making overlay/superposed stacks of names/text

\usepackage[final]{pdfcomment} % for tooltips
\usepackage{accsupp}           % for accessibility
\usepackage{pgf}               % for transparency

% Usage: \textopacity{opacity}{text}
% transparency.sty doesn't blend well in Acrobat, so we use PGF;
% see https://tex.stackexchange.com/a/681324/245104
\newcommand\textopacity[2]{%
  \begin{pgfpicture}%
    \pgfsetfillopacity{#1}%
    \pgfpathmoveto{\pgfpointorigin}%
    \pgftext[base]{#2}%
  \end{pgfpicture}%
}

% Usage: \textstack{opacity}{item1; item2; ...}
% Or: \textstack[sep]{opacity}{item1 sep item2 sep ...}
% Expands to \vbox{\hbox{item1}\vskip-\baselineskip\hbox{item2}...},
% plus a PDF tooltip, copy/paste override, and specified opacity.
% Loop iteration based on https://tex.stackexchange.com/a/159177/245104
\makeatletter
\newcommand\textstack[3][;]{%
  % Append separator (#1) and \@eol to end of input:
  \def\textstack@append##1{\expandafter\textstack@step##1#1\@eol}%
  % Split into first component (##1) up to separator (#1), and rest (##2):
  \def\textstack@step##1#1##2\@eol{%
    % Build \hbox for this component (##1) with specified opacity (#2)
    \hbox{\textopacity{#2}{\ignorespaces ##1\unskip}}%
    \ifx\@eol##2\@eol\else
      % More steps: unwind vertical space and continue
      \vskip-\baselineskip
      \textstack@step##2\@eol
    \fi}%
  % Tooltip with full author list (#3)
  \pdftooltip{%
    % Override copy/paste text with full author list (#3)
    \BeginAccSupp{method=pdfstringdef,unicode,ActualText={#3}}%
      % Wrap stack in \vbox
      \vbox{\textstack@append{#3}}%
    \EndAccSupp{}%
  }{#3}%
}
\makeatother

% Usage: \namestack{name1; name2; ...}
% Or: \namestack[opacity]{name1; name2; ...}
% Like \textstack but with optional opacity defaulting to 2/3
\newcommand\namestack[2][0.666]{\textstack{#1}{#2}}
\end{lstlisting}

\begin{lstlisting}[float=*,caption={\LaTeX\ code providing additional support for \texttt{natbib.sty}},label=lst:natbib]
%% \namestack support for natbib.sty
\makeatletter

% When formatting the name part of a citation, wrap in \namestack
\def\NAT@nmfmt#1{%
  \xdef\textstack@expand{#1}%
  \expandafter\namestack\textstack@expand
}

% Remove links from names so that tooltips are visible.
% Code from https://tex.stackexchange.com/a/27311/2451040
\usepackage{etoolbox}

% Patch case where name and year are separated by aysep
\patchcmd{\NAT@citex}
  {\@citea\NAT@hyper@{%
     \NAT@nmfmt{\NAT@nm}%
     \hyper@natlinkbreak{\NAT@aysep\NAT@spacechar}{\@citeb\@extra@b@citeb}%
     \NAT@date}}
  {\@citea\NAT@nmfmt{\NAT@nm}%
   \NAT@aysep\NAT@spacechar\NAT@hyper@{\NAT@date}}{}{}

% Patch case where name and year are separated by opening bracket
\patchcmd{\NAT@citex}
  {\@citea\NAT@hyper@{%
     \NAT@nmfmt{\NAT@nm}%
     \hyper@natlinkbreak{\NAT@spacechar\NAT@@open\if*#1*\else#1\NAT@spacechar\fi}%
       {\@citeb\@extra@b@citeb}%
     \NAT@date}}
  {\@citea\NAT@nmfmt{\NAT@nm}%
   \NAT@spacechar\NAT@@open\if*#1*\else#1\NAT@spacechar\fi\NAT@hyper@{\NAT@date}}
  {}{}

\makeatother
\end{lstlisting}

\lstset{
  alsoletter={$,.},
  keywords={FUNCTION, if$, while$, for$, swap$, num.names$, format.name$, int.to.str$, write$, newline$, cite$},
  string=[b]",
  emph={namestack, format.names, format.full.names, make.full.names, calc.label, write.label, output.bibitem, before.all, output.state},
}

\begin{lstlisting}[float=*,caption={Excerpts from \BibTeX\ style file, including all changed functions from \texttt{abbrvnat.bst}},label=lst:bst,multicols=2]
%% File: `stack-abbrvnat.bst'
%% A modification of `abbrv.bst' for use
%% with natbib package, further modified
%% for stacking names with \namestack

% Wrap top of stack in \namestack{...}
FUNCTION {namestack}
{ "\namestack{" swap$ * "}" *
}

% Format list of names (author or editor)
% on top of stack for inclusion in bibitem
FUNCTION {format.names}
{ 's :=
  #1 'nameptr :=
  s num.names$ 'numnames :=
  numnames 'namesleft :=
    { namesleft #0 > }
    { s nameptr "{f.~}{vv~}{ll}{, jj}"
      format.name$ 't :=
      nameptr #1 >
        { namesleft #1 >
            { "; " * t * }
            { t "others" =
                { " et~al." * }
                { "; " * t * }
              if$
            }
          if$
        }
        't
      if$
      nameptr #1 + 'nameptr :=
      namesleft #1 - 'namesleft :=
    }
  while$
  namestack
  % Append parenthetical number of names
  " (" * numnames int.to.str$ * ")" *
}

% Format list of names (author or editor)
% on top of stack for inclusion in
% natbib.sty label (just last names)
FUNCTION {format.full.names}
{'s :=
  #1 'nameptr :=
  s num.names$ 'numnames :=
  numnames 'namesleft :=
    { namesleft #0 > }
    { s nameptr
      "{vv }{ll}" format.name$ 't :=
      nameptr #1 >
        { "; " * t * }
        't
      if$
      nameptr #1 + 'nameptr :=
      namesleft #1 - 'namesleft :=
    }
  while$
}

% Omitted because not changed: make.full.names
% calls format.full.names with author or editor

% Compute label for sorting purposes (not used
% for actual output because it can overflow)
FUNCTION {calc.label}
{
  make.full.names
  "(" *
  year *
  ")" *
  'label :=
}

% Output natbib.sty label, without storing in
% a variable so won't overflow with many names
FUNCTION {write.label}
{
  make.full.names write$
  "(" write$
  year write$
  ")" write$
}

FUNCTION {output.bibitem}
{ newline$
  "\bibitem[" write$
  write.label
  "]{" write$
  cite$ write$
  "}" write$
  newline$
  ""
  before.all 'output.state :=
}
\end{lstlisting}

\bibliography{paper}
\bibliographystyle{stack-abbrvnat}

\end{document}